\documentclass[aps,twocolumn,prl,floatfix]{revtex4}
\usepackage{amsmath}
\usepackage{graphicx}
\usepackage{amsfonts}
\usepackage{amssymb}
\input epsf
\def\Z{{\mathbb Z}}

\begin{document}

\title{Lack of consensus in social systems}
\author{I.J. Benczik, S.Z. Benczik, B. Schmittmann, and R.K.P. Zia}
\affiliation{Department of Physics, Virginia Tech, Blacksburg, VA 24061-0435
USA}

\date{\today }
\begin{abstract}
We propose an exactly solvable model for the dynamics of voters in a two-party system. The opinion 
formation process is modeled on a random network of agents. The dynamical nature of interpersonal 
relations is also reflected in the model, as the connections in the network evolve with the 
dynamics of the voters. In the infinite time limit, an exact solution predicts the emergence of 
consensus, for arbitrary initial conditions. However, before consensus is reached, two different 
metastable states can persist for exponentially long times. One state reflects a perfect balancing 
of opinions, the other reflects a completely static situation. An estimate of the associated 
lifetimes suggests that lack of consensus is typical for large systems. 
\end{abstract}

\maketitle

Concepts and tools of modern nonequilibrium statistical physics lend themselves very directly to 
describing complex interacting systems, including phenomena which rely on human behavior, e.g. the 
emergence of collective organization in social systems. Recently variants of the voter model \cite
{FK96,S70,K92} have been used intensively to study collective phenomena, such as opinion formation 
or consensus creation \cite{CMV00,D01,M03,VKR03,K03,TTA00,DV02,SL03,G02,CVV03,SES05,VC04}. Many of
these efforts have focused on regular lattices \cite{CMV00,D01,M03,VKR03,K03,TTA00}, which is 
justified in physical situations, but not in the context of the social sciences. In socio-cultural 
situations, the interaction patterns between individuals find a better characterization as complex 
networks in which the connections or relationships (links) between individuals (nodes) can change in 
time.  More precisely, the full dynamics of such a social network consists of (i) the opinion 
formation process taking place on the nodes, and (ii) the evolution of the underlying topological 
structure (links). The coupling between these two processes reflects how the connections of people 
influence their opinions, and how their opinions determine, in turn, their new connections. Although 
there is an increasing recent interest in modeling voter dynamics on graphs \cite{DV02,SL03}, and on 
networks \cite{G02,SES05,CVV03}, the dynamics of the links (ii) is still disregarded in these studies.
The coevolution of nodes \emph{and} links -- i.e., of the full network structure -- has been studied 
only in a few works \cite{GZ06,VC04,VR07,LB06,AP06,HN06}.

In this letter, we investigate the voter dynamics on an \emph{adaptive} disordered network characterized
by (i)-(ii). In our network, each node $j$ (``individual'') carries a spin $\sigma _j$ (``opinion'')
which can take two different values $\sigma _j=\pm 1$ \cite{S70}. At each time step, (i) the spins are 
updated random sequentially based on a simple majority rule: if they are connected to more positive than
negative spins, their state will be positive in the next time step, and negative otherwise; in the case 
of a tie, the spin remains unchanged. Further, (ii) the links are updated as follows: two nodes carrying equal
(unequal) spins are connected with probability $p$ ($q$). In this letter, we focus on the special case 
$q=1-p$, leaving the general case to \cite{promises}. 

As an interpretation, we propose that this model mimics a two-party electoral system. During a campaign, 
the supporters of one party are keen to interact with supporters of the other party 
to try to change their opinion. This situation can be described by this model 
with $p<q$, when each agent has more interactions with opponents than with agents sharing the same opinion
(according to the motto that \emph{``convinced people do not need to be convinced again''}). On the other
hand, when $p>q$, the agents tend to interact more with individuals sharing the same opinion
(according to the motto \emph{``united we are stronger''}). The latter behavior seems to be a simplified
description of the process of political \emph{polarization}, when all the members of a party agree 
with the official position of the party, as often occurs in post-election periods.

Our model is aimed to describe a free public debate in the sense that it does not consider the effects 
of central institutions or\ the mass media; neither lobbying, nor organized strategies (apart from 
possibly influencing the probability $p$) are taken into account. As a result, the model may also
be appropriate to describe groups defined by criteria such as education, religion or ethnicity, rather
than political opinion. Cultural assimilation, the spreading of a language or a religion of an ethnic 
or religious minority, and social reforms are examples of phenomena which 
can be modeled in this fashion.

To describe the dynamics of the system, let us focus on $\rho \left(t\right) $, the ``popularity'' of 
$+$ opinions, defined as the average fraction of $+$ nodes at time $t$. Thus, we consider $P(M,t)$, 
the probability of finding the network with $M$ positive spins at time $t$, and 
\begin{equation}
\rho (t)\equiv \sum_{M=0}^N\frac{M}NP(M,t),
\end{equation}
where $N$ is the total (fixed)\ number of nodes. Clearly, $\rho =0$ or $1$ correspond to a complete 
ordering of the system, while $\rho =0.5$ characterizes the completely disordered state. Contrary to 
the voter model on regular lattices, the global magnetization ($m=2 \rho-1$) is \emph{not conserved} 
here, but the dynamics is still $\Z_2$ symmetric (i.e., invariant under the global inversion $\sigma 
_i\mapsto - \sigma _i$, $M/N\mapsto 1-M/N$). 

Since the spins on the nodes flip one at a time, the time evolution of $P(M,t)$ is a simple birth-death
process for which we can write a master equation: 
\begin{align} \begin{split}
\partial _{t}P(M,t)& =b_{M-1}P(M-1,t)+d_{M+1}P(M+1,t) \\
& \quad -\left[ b_{M}+d_{M}\right] P\left( M,t\right). \label{master}
\end{split} \end{align}
Here, $b_{M}$ denotes the birth rate of a positive spin (i.e., the rate for flipping a negative to a
positive spin), and $d_{M}$ its death rate (flip rate from positive to negative). Both depend on $M$, 
the current number of positive spins in the system. Whether a positive spin will flip or not is 
determined by the number of positive and negative spins it is connected to. Since these connections 
are established randomly, with probabilities $p$ and $q$, respectively, they are controlled by binomial
distributions. For example, the probability that a positive spin is connected to exactly $k$ of the other
$M-1$ positive spins, is given by $B_{M-1,\,p}(k)\equiv \binom{M-1}{k}p^{k}(1-p)^{M-1-k}$. Writing a 
similar expression for the probability of this spin to be connected to $k^{\prime }$ negative spins, 
the death rate is given by
\[d_M=\frac{M}{N}\sum_{k=0}^{M-1}\sum_{k^\prime =0}^{N-M} B_{M-1,\,p}(k)\,B_{N-M,\,q}(k^{\prime })\,
\Theta (k^{\prime }-k).\]
The prefactor simply reflects the probability to find a positive spin among the $N$ spins. The step 
function $\Theta (k^{\prime }-k)$ expresses the fact that if the selected spin is connected to $k$ 
positive and $k^{\prime }$ negative spins then, in the next time step, it will take the state $
\mathop{\text{sgn}}\left( k-k^{\prime }\right)$. Similarly, the birth rate is: 
\[b_{M}=\frac{N-M}{N}\sum_{l=0}^{N-M-1}\sum_{l^{\prime }=0}^{M}B_{N-M-1,p}(l)
B_{M,q}(l^{\prime }) \Theta (l^\prime-l).\]
For the special case $q=1-p$ \cite{promises}, the $\Theta $-functions can be eliminated from $b_{M}$ 
and $d_{M}$, due to the properties of the binomial distributions, to yield:
\begin{align}
d_{M}& =\frac{M}{N}\sum_{k=0}^{N-M-1}B_{N-1,p}(k), \\
b_{M}& =\frac{N-M}{N}\sum_{l=0}^{M-1}B_{N-1,p}(l).
\end{align}
\begin{figure}[htbp]
\epsfxsize=3.4 in \centerline{\epsfbox{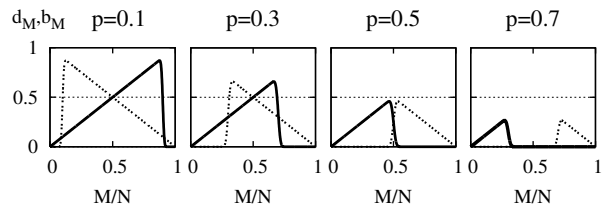}}
\caption{The death $d_M$ and birth $d_M$ rates (continuous and dashed lines, respectively), for $N=1000$
and different probabilities $p$. For fixed $p$, there are three different regimes of behavior bounded 
by Eq.~(\ref{bound}).} \label{cf}
\end{figure}

Before solving the master equation, we discuss its approximate solutions in the thermodynamic 
limit, in order to identify the different types of behavior and the parameter regimes where they occur. 
For $N\rightarrow \infty $, the binomial distribution $B_{N,p}(k)$ approaches a normal distribution with
mean $Np$, so that $d_{M}$ is given by the Gaussian error function 
multiplied by the prefactor $M/N$, 
\begin{equation}
d_{M}\simeq \begin{cases} M/N, & \text{if }M<Np, \\ 0, & \text{if }M>Np,\end{cases}  \label{approx}
\end{equation}
apart from a region of width $\sqrt{Np(1-p)}$ around $Np$. Similarly, the birth rate $b_{M}$ is described
by the complementary error function.
These forms of the transition rates (see Fig.~\ref{cf}), along with the probability $p$ 
and the initial fraction of positive spins, $M_0/N$, determine the 
late-time properties of the model. The master equation controls the flow of $M/N$, as a function of time, 
leading to four distinct regimes, depending on the relative magnitudes of $M_0/N$\ and $p$. In a 
$(p,M_0/N)$ phase diagram, these different regimes are bounded by 
\begin{equation}
\frac{M_0}N=p,\quad \mbox{and}\quad \frac{M_0}N=1-p. \label{bound}
\end{equation}

For $p<0.5$ and $M_0/N<p$, we find that $M/N$ stays below $p$ at later time also. Indeed, in the 
approximation (\ref{approx}), we have a pure death process 
\begin{equation}
\partial _tP(M,t)=\left( M+1\right) P(M+1,t)-MP\left( M,t\right), \label{master1aa}
\end{equation}
which leads to the extinction of the positive population. The steady state, 
$\rho _\infty \equiv \lim_{t\rightarrow \infty }\rho (t)=0$, is reached
exponentially as $\rho (t)\sim \rho _0\exp (-t/N)$. Similarly, if $p<0.5$
and $M_0/N>1-p$, we have a pure birth process, and the system relaxes
exponentially to the state $\rho _\infty =1$ on the same characteristic time
scale as in the previous case (due to the $\mathbb{Z}_2$-symmetry). In the
intermediate region $M_0/N\in [p,1-p]$, the dynamics is described by:
\begin{align}
\nonumber
N \partial _tP(M,&t)=( M+1) P(M+1,t) \\ +&(N-M+1) P(M-1,t) - N P(M,t),
\end{align}
and the system reaches a disordered phase: $\rho _\infty =0.5$. Again, the relaxation is exponential, 
with a characteristic time scale $N/2$. 

For $p>0.5$ the pure death and pure birth regimes are the same as for $p<0.5$. A small minority ($M_{0}/
N<1-p$) will become extinct, a large majority ($M_{0}/N>p$) will win. However, a new feature appears in 
the interval $M_{0}/N\in [1-p,p]$, where the system seems to acquire infinite memory. Both the death and
birth rates vanish in this region, so that $\partial _{t}
P(M,t)=0$ whence the fraction of positive spins remains frozen at its initial value, $M_{0}/N$. 

\begin{figure}[htbp]
\epsfxsize=3.4 in \centerline{\epsfbox{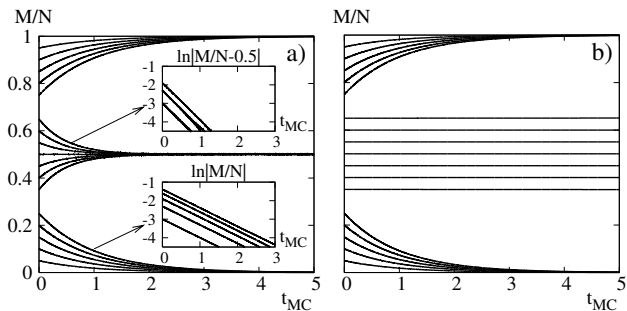}}
\caption{Time evolution of the fraction $M/N$ of positive spins for a network of $N=1000$ nodes, for $p=
0.3$ (a) and $p=0.7$ (b). The values are averaged over $1000$ runs. Time is measured in Monte Carlo steps,
$t_{\text{MC}} =t_{\text{spin-flip}}/N$. The insets show the exponential relaxation to the final states.} 
\label{fig3}
\end{figure}

\begin{figure}[htbp]
\epsfxsize=3.6 in \centerline{\epsfbox{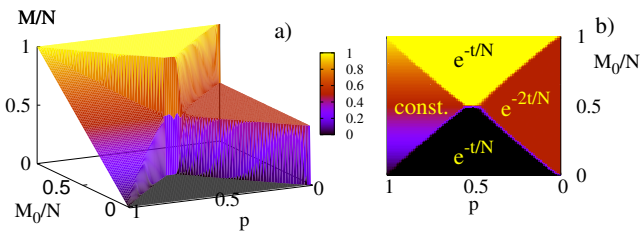}} \vspace*{-0.7cm}
\caption{The fraction of positive spins $M/N$ after 
$t_{\text{MC}} =10$, as function of $p$ and $M_{0}/N$ for a network of $N=1,000$ nodes in $3D$ (a), and 
$2D$ (b) displaying clearly the phase transitions. 
} \label{fig4}
\end{figure}

Our analytic findings are tested by simulations \cite{KS88} on a network with 
$N=1,000$ nodes. The relaxation into the four late-time states is displayed 
in Figs.~\ref{fig3}(a) for $p<0.5$ and (b) for $p>0.5$. The possible outcomes of the
voter dynamics, for all parameters $p$ and initial
fractions $M_{0}/N$ of the positive population are summarized in Fig.~\ref{fig4}. 

\begin{figure}[htbp]
\epsfxsize=2.4 in \centerline{\epsfbox{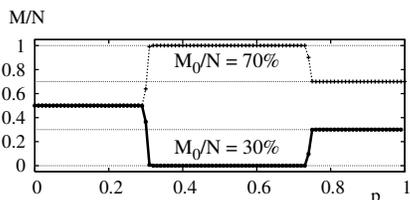}} \vspace*{-0.4cm}
\caption{Outcome of the voter dynamics as function of $p$. Cross sections of
Fig.~\ref{fig4} taken along $M_{0}/N=0.3$ and $M_{0}/N=0.7$. } \label{fig4a}
\end{figure}

To illustrate the picture further, Fig.~\ref{fig4a} shows the outcome of the voter dynamics for two
initial fractions of positive population: one starting from a minority $M_0/N<0.5$, and one starting from 
a majority $M_0/N>0.5$. For small $p$, the system reaches a disordered state, independent of
$M_0/N$: the ``open mindedness'' of the population (reflected by a large probability $1-p$ to communicate 
with the opposite party) leads to an equal distribution of opinions. In contrast, an ``inflexible attitude'' 
(characterized by a large probability $p$ of linking up with similar opinions) leads to an unchanging
distribution of opinions. For intermediate values of $p$, the system reaches a completely ordered state: 
all voters reach the same opinion, namely that of the initial majority. 

In conclusion, in the thermodynamic limit, the voter dynamics has four possible outcomes: a perfect balance
of opinions, a static situation, or consensus ($\rho =0,1$). In this last section, we 
discuss how these findings are modified in \emph {finite} systems. The two completely ordered 
states  are absorbing states, thus they will be reached from the other two 
(metastable) states. Two interesting questions remain: First, in \emph {which} of the two absorbing states
will each metastable state arrive, and second, how do the relaxation times depend on system size? 

To answer the first question, we write Eq.~\eqref{master} in a matrix representation, 
$\partial_t\left| v(t)\right\rangle =\mathbb{L}\left| v(t)\right\rangle$, where $\left| v(t)\right\rangle$
is the $\left( N+1\right) $-dimensional column vector with components $P(M,t)$, $M=0,1, ...,N$, and the 
time evolution operator $\mathbb{L}$ can be read off from Eq.~\eqref{master}. The steady
states of the system are the eigenvectors of  $\mathbb{L}$ with zero eigenvalues: 
$\lvert 0_0\rangle \equiv 
\begin{pmatrix} 
1, & 0, & ...,  & 0%
\end{pmatrix}^{\mathsf{T}}$ and $\lvert 0_N\rangle \equiv 
\begin{pmatrix} 
0, & 0, & ...,  & 1 
\end{pmatrix}^{\mathsf{T}}$. To find their adjoints, it is convenient to study the symmetric/antisymmetric 
states: $\lvert 0_{\pm }\rangle \equiv \big(\lvert 0_0\rangle \pm \lvert 0_N\rangle \big)/2$. Imposing 
orthonormality $\langle 0_{\pm }|0_{\pm }\rangle =\delta _{\pm }$, we obtain from $\langle 0_{\pm }\rvert
\,\mathbb{L}=0$ the right eigenvectors $\langle 0_{+} \rvert \equiv 
\begin{pmatrix} 
1, & 1, & ..., & 1 
\end{pmatrix}$ and $\langle 0_{-}\rvert \equiv 
\begin{pmatrix} 
1, & x_{1}, & x_{2}, & ..., & -x_{2}, & -x_{1}, & -1 
\end{pmatrix}$, with 
\begin{equation}
x_j\equiv \frac{\big(((r_{j+1}+1)r_{j+2}+1\big)r_{j+3}+... +1\big)r_{n-1}+1}{\big(((r_1+1)r_2+1\big)r_3+... +1
\big)r_{n-1}+1}, 
\nonumber
\end{equation}
where $r_j\equiv b_j/d_j$. Thus we can compute explicitly the final state $\lvert \psi _\infty 
\rangle \equiv \lim_{t\rightarrow \infty }\lvert \psi _t\rangle$. Given an initial state $\lvert \psi _0
\rangle $, the solution to Eq.~\eqref{master} is 
$\big\lvert\psi _t\big\rangle=\sum_\mu e^{-\lambda _\mu t}\lvert \mu \rangle
\langle \mu |\psi _0\rangle $,
where $\langle \mu \rvert $ and $\lvert \mu \rangle $ are the left and right eigenvectors of $\mathbb{L}$ 
corresponding to eigenvalues $\lambda _\mu $. Expecting no other zero $\lambda _\mu $'s, we have $\lvert 
\psi _\infty \rangle =\lvert 0_{+}\rangle \langle 0_{+}|\psi _0\rangle +\lvert 0_{-}\rangle \langle 0_{-}|
\psi _0\rangle $. For example, if the initial state is a population with $M_0<N/2$ positive spins, the 
final state is 
\begin{equation}
\lvert \psi _\infty \rangle = \frac 12\begin{pmatrix} 1{+}x_{M_{0}}, & 0, & ..., & 0, &
1{-}x_{M_{0}}\end{pmatrix}^{\mathsf{T}}. \label{result}
\end{equation}
This result shows that, indeed, in the $t \to \infty $ limit, the system arrives in one of its two 
absorbing states. Moreover, it provides the \emph{relative probabilities} with which either will be 
reached, as a function of the initial $M_0$. Further details will be published elsewhere \cite{promises}. 

\begin{figure}[tbph]
\epsfxsize=3.0 in \centerline{\epsfbox{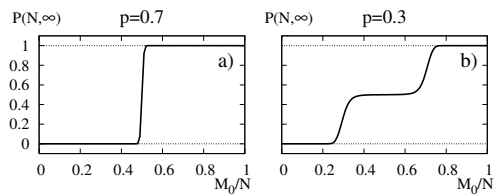}} \vspace*{0cm}
\caption{The probability $P(N,\infty )=(1-x_M)/2$, to reach the state with 
$M=N$ positive spins at infinite times: $P(N,\infty )=0$ implies certain
extinction of the positive population, while $P(N,\infty )=1$ represents a
purely positive population.}
\label{fig6}
\vspace{-0.3cm}
\end{figure}

These findings are confirmed in Fig.~\ref{fig6}, obtained by direct iteration of the master equation. For 
$p>0.5$, the dynamics reduces to a simple majority rule: The final state is completely ordered, following 
the opinion of the initial majority. For $p<0.5$, a small minority disappears, while a big majority wins 
the competition, just as in the case of infinite system size. A novel effect occurs for \emph{intermediate}
values of the initial positive population. In \emph{finite} systems, the fully disordered (metastable) 
state has a finite lifetime, during which the system forgets its initial condition. Then, after a very long
time, it \emph{randomly} falls into one of its absorbing states. To rephrase, the initial positive 
population is equally likely to become extinct or to take over the whole system. In this regime, the final 
outcome of the voter dynamics is completely random.

\begin{figure}[tbp]
\epsfxsize=3.2 in \centerline{\epsfbox{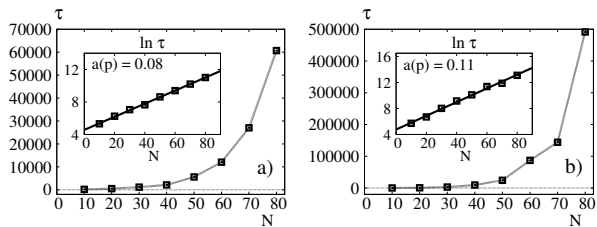}} 
\caption{Time $\tau$ needed to reach consensus as function of the system size for $p=0.3, M_0/N=0.4$ (a), 
and $p=0.8, M_0/N=0.4$ (b). The insets show the exponential dependence on $N$. } \label{fig7}
\end{figure}

Finally, we explore numerically the relaxation times into the absorbing states in Fig.~\ref{fig7}. The 
lifetime $\tau$ of the metastable states increases with the system size $N$, as $\tau \sim e^{a (p)
N}$, with a $p$-dependent coefficient $a (p)$. An estimate of the time to consensus for a network of 
$N=1000$ voters with $p=0.3$ shows that it may take as many as $10^{36}$ spin flips to reach one of the 
absorbing states. In much larger systems, consensus is practically impossible.

In conclusion, the best strategy for a minority group is to establish many contacts with 
its opponents. In this way, it can convince half of them and keep this balance for a long time. If the same
group is less open for discussions, it cannot overcome the majority, but at least it will not disappear. 
It is tempting to speculate how these results might be applied to real social systems.
Will two-party systems, once formed, persist for long times? Will bilingual regions remain 
bilingual? Will relatively isolated parties continue to receive the same, almost constant percentage 
of the vote? Will closed religious communities continue to exist without gaining or losing members?

These particular results are obtained from a simple adaptive model in which certain important social 
factors are neglected (e.g., spatial and age structures, a spectrum of opinions, etc.). More precise 
statements will certainly need additional assumptions on the character of the network or opinion dynamics. 
Nevertheless, even this simple model can give a 
better understanding of empirical data not yet explained. For example, a study of the number of languages
in the Solomon Islands \cite{T77} found that small islands (less than 100 square miles) have a single 
language, but above this size the number of languages increases. The finite size effect pointed out by
our model can be a possible explanation of this phenomenon.

The importance of the model presented here stems from the fact that it is mathematically simple, 
exactly solvable, and easily generalized to more complex situations (for $q\neq 1-p$ see \cite{promises}, 
for epidemics networks see \cite{promises2}). Our model provides a \emph{method} how to describe 
analytically the adaptive nature of the interpersonal relations, and by this, it can
serve as a ``baseline model'' which captures the key characteristics of social systems, namely, having 
disordered networks of agents and dynamically changing connections between them. 

We thank S. Redner, M. Mobilia and I. Georgiev for useful discussions, and K. Holsinger for the binomial
generator. This research is supported by NSF through DMR-0414122 and
the College of Science at Virginia Tech.

\end{document}